\documentclass[]{aa}
\usepackage[affil-it]{authblk}
\usepackage{graphicx}
\usepackage[space]{grffile}
\usepackage{latexsym}
\usepackage{amsfonts,amsmath,amssymb}
\usepackage{url}
\usepackage[utf8]{inputenc}
\usepackage{hyperref}
\hypersetup{colorlinks=false,pdfborder={0 0 0}}
\usepackage{textcomp}
\usepackage{longtable}
\usepackage{natbib}
\usepackage{multirow,booktabs}
\usepackage{auto-pst-pdf}
\usepackage{color}
\usepackage{gensymb} 

\begin{document}

\title{OmegaWINGS: spectroscopy in the outskirts of local clusters of galaxies
}

   \author{
A. Moretti      \inst{1}\and
M. Gullieuszik  \inst{1}\and
B. Poggianti    \inst{1}\and
A. Paccagnella  \inst{1,2}\and
W. J. Couch     \inst{3}\and
B. Vulcani      \inst{4}\and
D. Bettoni      \inst{1}\and
J. Fritz        \inst{5}\and
A. Cava         \inst{6}\and
G. Fasano       \inst{1}\and
M. D'Onofrio    \inst{1,2}\and
A. Omizzolo \inst{1,7}
   }
   \institute{
     INAF - Osservatorio astronomico di Padova, Vicolo
     dell'Osservatorio 5, 35122 Padova, Italy
     \and
     Dipartimento di Fisica e Astronomia, Universit\`a degli Studi di
     Padova, Vicolo dell'Osservatorio 3, 35122 Padova, Italy
     \and
     Australian Astronomical Observatory, PO Box 915, North Ryde, NSW
     1670 Australia
     \and
     School of Physics, University of Melbourne, VIC 3010, Australia
     \and
     Instituto de Radioastronom\'\i a y Astrof\'\i sica, UNAM, Campus Morelia, A.P. 3-72, C.P. 58089, Mexico
     \and
     Observatoire de Gen\`eve, Universit\`e de Gen\`eve, 51 Ch. des
     Maillettes, 1290 Versoix, Switzerland
     \and
     Specola Vaticana, 00120, Vatican City State
}

\date{\today}
\abstract 
{Studies of the properties of low-redshift cluster galaxies suffer, in general, from small spatial coverage of the cluster area. WINGS, the most homogeneous and complete study of galaxies in dense environments to date, obtained spectroscopic redshifts for 48 clusters at a median redshift of 0.05, out to an average distance of approximately $0.5$ cluster virial radii. The WINGS photometric survey was recently extended by the VST survey OmegaWINGS to cover the outskirts of a subset of the original cluster sample.}
{In this work, we present the spectroscopic follow-up of 33 of the 46 clusters of galaxies observed with VST over 1 square degree. The aim of this spectroscopic survey is to enlarge the number of cluster members and study the galaxy characteristics and the cluster dynamical properties out to large radii, reaching the virial radius and beyond.}
{We used the AAOmega spectrograph at AAT to obtain fiber-integrated spectra covering the wavelength region between $3800$ and $9000$\, \AA\, with a spectral resolution of $3.5-6$\,\AA\,  full width at half maximum (FWHM). Observations were performed using two different configurations and exposure times per cluster. We measured redshifts using both absorption and emission lines and used them to derive the cluster redshifts and velocity dispersions.}
{We present here the redshift measurements for 17985 galaxies, 7497 of which turned out to be cluster members. The sample magnitude completeness is $80\%$ at V=20.
Thanks to the observing strategy, the radial completeness turned out to be relatively constant ($90\%$) within the AAOmega field of view. The success rate in measuring redshifts is $95\%$, at all radii.}
{We provide redshifts for the full sample of galaxies in OmegaWINGS clusters together with updated and robust cluster redshift and velocity dispersions. These data, publicly accessible through the CDS and VO archives, will enable evolutionary and environmental studies of cluster properties, providing a local benchmark.
}
\authorrunning{Moretti et al.}
\titlerunning{OmegaWINGS spectroscopy}
\maketitle 

\section{Introduction}
The WINGS survey \citep{Fasano+2006} was the first attempt to map local galaxy clusters (median redshift $\sim 0.05$, 76 clusters) over a wide field of view in the optical range. 
With respect to other surveys of nearby galaxies, such as the Sloan Digital Sky Survey (SDSS) \citep{sdss}, it offers the advantage of being targeted to X-ray-selected clusters. Moreover, it reaches galaxy magnitudes that are 1.5 mag deeper than the ones obtained by the widely used SDSS.

The WINGS survey was extended in the near IR \citep{Valentinuzzi2009} for 28 clusters using data collected with the WFCAM camera at the UKIRT, covering a region four times larger than the optical one.

The spectroscopic follow-up \citep{Cava2009} allowed the measurement of approximately $6000$ new redshifts in 48 out of the 76 clusters and the determination of accurate velocity dispersions and membership.
Only 21/48 clusters, though, had at least $50\%$ completeness and could be used in statistical studies \citep{Valentinuzzi+2010,Vulcani+2011b,vulcani+2011c,Vulcani2012,Moretti2015}.
The complete set of measurements made on the WINGS dataset is available using the Virtual Observatory tool \citep{Moretti2014}.

The WINGS coverage of $\sim 30 \times 30$ arcmin$^2$ often covers just half of the cluster virial radius, and is thus missing the more external regions of the clusters.
This represents a serious limitation for the comprehension of galaxy formation in clusters, as the cluster outskirts are the regions where galaxies probably undergo the major transformations that reflect, for instance, into the well known morphology--density relation \citep{dressler80}.
In particular, \citet{Fasano2015}, using the WINGS dataset, found that this relation seems to disappear outside of the cluster core, supporting the notion that
the relation between the morphological mix of galaxies and the distance from the cluster center might be more prominent than that with local density in these regions. Several other studies have pointed out the dependence of galaxy properties on cluster-centric radius also at large distances \citep{Lewis2002,Gomez2002}.

Moreover, studies on cluster luminosity functions (LF) highlighted how, when comparing LFs in different clusters, it is mandatory to compare the same physical regions \citep{Popesso06,Barkhouse07,Moretti2015}, which, so far, have always been  restricted to $0.5R_{200}$.

Cluster outskirts are the places where galaxies are either individually accreted, or processed in groups and filaments that eventually fall into the cluster potential, and therefore represent the transition region where galaxy transformations take place \citep{Lewis2002,Pimbblet2002,Moran2007}.

This specific transition region has been studied in detail in a few single clusters and superclusters \citep{Merluzzi2010,Haines2011,jaffe2011,Smith2012,Merluzzi2015},  but, up to now, a systematic study of cluster outskirts has been hampered by the lack of an appropriate set of observational data.

Very recently, the WINGS survey has been extended with OMEGACAM at VST \citep{Gullieuszik+2015}, with the aim of covering the virial radius of the observed clusters. The OmegaWINGS photometric survey covers 57 out of the original 76 clusters (among those with ($\delta \leq 20^{\degree}$), visible with the VST).

The OmegaWINGS sample of cluster galaxies has already been used to investigate properties of transition galaxies out to the virial radius and beyond and led to the discovery of a widespread population of galaxies with suppressed Specific Star Formation Rate (SSFR), possibly undergoing a slow quenching of star formation \citep{Paccagnella2016}.

This paper describes the spectroscopic follow-up of the OmegaWINGS survey.
Section \ref{sec:strategy} describes the sample and the target selection criteria, section \ref{sec:obs} describes the spectroscopic observations and data reduction, while in section \ref{sec:quality}, we assess the data quality, including the signal to noise and the completeness. In section \ref{sec:redshifts}, we illustrate the redshift measurement procedure, and in section \ref{sec:sigma}, we give the principal products of the spectroscopic catalog, that is, velocity dispersions and memberships.
Finally, in section \ref{sec:access}, we describe how to access the entire set of measurements. A short summary is given then in section \ref{sec:summary}.

\section{Sample and survey strategy}\label{sec:strategy}
The OmegaWINGS spectroscopic survey is based on the photometric observations of 46  clusters (out of the 57 original OmegaWINGS sample).

The VST images, both in the $V$ and $B$ bands, cover a region of approximately 1 deg$^2$ around the cluster center, as derived from the {\it SIMBAD} database \citep{simbad}, and are $50 \%$ complete at V$\sim 23$ mag.
More important for the spectroscopic survey purposes is the internal accuracy of the astrometric calibration, that is equal to or better than $0.1"$.

The selection of candidates for the spectroscopic follow up was made using the OmegaWINGS photometric catalogs by \citet{Gullieuszik+2015} on the basis of object magnitude and color.
We selected the detections classified as galaxies with a total
 (Sextractor AUTO) magnitude brighter than $V=20$ mag, and subsequently divided this sample into {\it bright} and {\it faint} sources according to their V aperture magnitude inside the fiber diameter ($2\farcs16$). Bright sources have $V_{fib}\leq20.5$ mag and faint sources have $20.5\le V_{fib}\le 21.5$. For each cluster, we therefore obtained two complementary sets of observations: one that has been exposed for 60 minutes, called {\it bright} configuration, and a second one, called {\it faint} configuration, with an exposure time of 120 minutes. 

To minimize the contamination from background galaxies, we applied a color cut to exclude extremely red background sources. This cut was set for each cluster by visually inspecting the color-magnitude diagram, in order to account for redshift variations, but in all cases, the color limit was very close to $B-V=1.20$ mag. 
Finally, we assigned a priority to the targets. The targets located outside the original WINGS area had the highest priority, followed by  the targets located  inside the WINGS images but without a spectrum (intermediate priority), and by the targets located inside the WINGS region and with a previously determined redshift (lowest priority).

\begin{figure}[h!]
\begin{center}
\includegraphics[width=0.33\textwidth,angle=90]{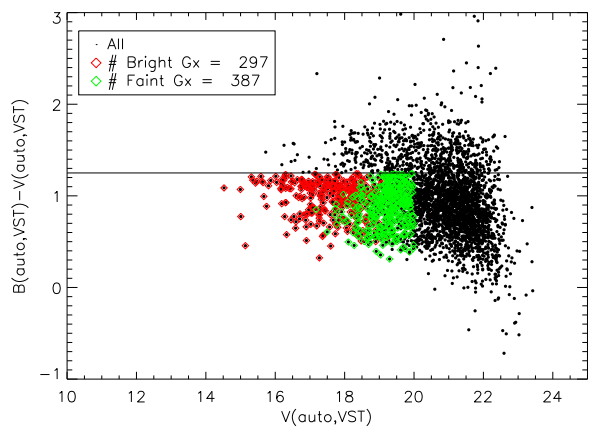}
\caption{Candidates for spectroscopic follow-up in A2382: the color cut is set at ($B-V$)=1.25. Black dots are sources flagged as galaxies and red and green symbols indicate galaxies belonging to the {\it bright} and {\it faint} configurations, respectively.}
\label{fig:selection}
\end{center}
\end{figure}

As an example, Fig. \ref{fig:selection} shows the target selection for the cluster A2382: the black dots are sources from \citet{Gullieuszik+2015} that have been classified as galaxies from the photometric survey, red points are galaxies belonging to the {\it bright}  configuration, while green dots are galaxies belonging to the {\it faint} configuration.

\section{Observations and data reduction}\label{sec:obs}
We used the AAOmega spectrograph \citep{Smith2004,Sharp2006} at the Australian Astronomical Observatory (AAT), which can host up to 392 fibers over $2\times2$ deg$^2$. 
The fiber diameter is $2.16\arcsec$.
The AAOmega spectrograph allows one to  simultaneously take the blue and the red spectra of a target galaxy, using different grisms. In particular, we used the 580V and the 385R grisms, which have a resolution $R=1300$ (FWHM=$3.5-6$ \AA ). The observed wavelength range is $\sim 3800-9000$ \AA.

For all configurations, we used 25 sky fibers, and 6 fibers for the guide stars, therefore we had $\sim 360$ available science 
fibers left per configuration.
The positions of the sky fibers were chosen by 
visually inspecting OmegaWINGS images and looking for 25 regions with no sources, uniformly distributed over the entire field of view.
Science targets were then positioned on the field by using the {\it configure} software, which takes into account the priorities described in Sect.\ref{sec:obs} using a Simulated Annealing (SA) algorithm \citep{aao_configure}.
A tumbling mechanism with two field plates allows the next field to be configured while the current field is being observed.

Observations started in August 2013 and were carried out in five different runs.
Table \ref{tab:log} contains the number of nights in the run, the number of observed targets for which we observed both configurations plus the number of configurations partially observed (either the {\it bright} or the {\it faint}), and the total success rate (i.e., the rate of successful redshift measurements over the total of assigned fibers in the run).
\begin{table}[htbp]
\label{tab:log}
\caption{Observations}
\begin{center}
\begin{tabular}{l||c|c|c}
\hline
Run & Nights & Targets & Success rate \\ 
\hline
August 2013       & 6 & 10+2  & 97.2 \% \\
December 2013     & 3 &  5+3  & 89.0 \% \\
July 2014         & 2 &  2+2  & 97.4 \% \\
January 2015      & 5 &  4+4  & 92.8 \% \\
September 2015    & 2 & 3+5   & 95.6 \% \\
\hline
\end{tabular}
\tablefoot{Number of observing nights, number of observed targets with both {\it faint} and {\it bright} configurations plus the number of configurations not completed, and average success rate of redshift measurements for each observational run.}
\end{center}
\end{table}%

Some observations were carried out under non-optimal weather conditions; in some cases we therefore decided to increase the total exposure time to reach a sufficient signal-to-noise ratio for all targets. 
The detailed observing log for each configuration is shown in Table \ref{tab:configurations} where we also report the seeing and the percentage of observed galaxies for which we could obtain reliable redshift measurements (i.e., success rate in the last column).
Some faint configurations (A168F, A3530F, A754F) could be exposed only for 1 hour, therefore obtaining spectra with a low S/N ($<5$), which translates to a lower success rate.
Arc exposures were taken before and after the scientific exposures.
Moreover, for the faint configurations, we always  also observed arc frames in between the science exposures, in order to properly calibrate in wavelength.
The cluster A3266 has a second, deeper faint configuration that has been exposed for 18000 s.
With these observations, we enlarged the number of WINGS clusters with spectroscopic follow-up, adding the clusters \object{A85}, \object{A168}, \object{A2717}, \object{A2734}, \object{A3532}, \object{A3528}, \object{A3530}, \object{A3558}, \object{A3667}, \object{A3716}, \object{A3880}, and \object{A4059}.

  \begin{table} 
  \caption{Observations log}
  \footnotesize
    \begin{tabular}{ l c c c c}
    \hline
Cluster/Conf. & Run & Seeing & Exp. time & Success rate \\
                      &         &  [arcsec]           & [s]          & \\ 
\hline
\object{A1069} B  & Dec13 & 2.0 & 3600 &  95    \\   
\object{A1069} F  & Dec13 & 1.8 & 6600 &  92    \\  
\object{A151} B   & Jan15 & 1.5 & 3600 &  96\\
\object{A151} F   & Jan15 & 1.8 & 7200 &  98 \\
\object{A1631a} B & Jul14 & 1.5 & 3600 &  99    \\    
\object{A1631a} F & Jul14 & 1.6 & 7200 &  93    \\    
\object{A168} B   & Dec13 & 1.8 & 3600 &  79   \\    
\object{A168} F   & Dec13 & 1.5 & 3600 &  77    \\    
\object{A193} B   & Dec13 & 1.6 & 3600 &  100   \\    
\object{A193} F   & Sep15 & 1.9 & 6000 & 100\\
\object{A2382} B  & Aug13 & 1.5 & 3600 &  100   \\    
\object{A2382} F  & Aug13 & 1.3 & 7200 &  97    \\    
\object{A2399} B  & Aug13 & 1.3 & 3600 &  97   \\   
\object{A2399} F  & Aug13 & 1.5 & 7200 &  91   \\    
\object{A2415} B  & Sep15 & 2.1 & 3600 & 92\\
\object{A2415} F  & Sep15 & 1.7 & 7200 & 99\\
\object{A2457} B  & Aug13 & 4.2 & 3600 &  99     \\   
\object{A2457} F  & Aug13 & 3.0 & 7200 &  100  \\   
\object{A2717} B  & Sep15 & 1.4 & 3600 & 99\\
\object{A2717} F  & Jul14 & 1.3 & 7200 &  94     \\   
\object{A2734} B  & Sep15 & 2.4 & 3600 & 87\\
\object{A2734} F  & Sep15 & 1.5 & 7200 & 93\\
\object{A3128} B  & Aug13 & 1.1 & 3600 &  100    \\   
\object{A3128} F  & Aug13 & 1.4 & 7200 &  99     \\   
\object{A3158} B  & Aug13 & 1.5 & 3600 &  100   \\   
\object{A3158} F  & Aug13 & 1.3 & 7200 &  99     \\   
\object{A3266} B  & Aug13 & 1.3 & 3600 &  100    \\   
\object{A3266} F  & Aug13 & 3.5 & 7200 &  89     \\   
\object{A3266} F* & Jan15 & 1.8 & 18000 & 82\\
\object{A3376} B  & Jan15 & 3.2 & 3600 &  92  \\
\object{A3376} F  & Jan15 & 1.6 & 7200 &  98 \\
\object{A3395} B  & Dec13 & 1.6 & 3600 &  96    \\   
\object{A3395} F  & Dec13 & 1.6 & 7200 &  97     \\   
\object{A3528} B  & Jan15 & 2.2 & 3600 &  99\\
\object{A3528} F  & Jan15 & 1.7 & 7200 &  86\\
\object{A3530} B  & Dec13 & 2.5 & 3600 &  96   \\    
\object{A3530} F  & Dec13 & 2.2 & 3600 &  67    \\      
\object{A3532} F  & Jan15 & 1.5 & 7200 &  88\\
\object{A3556} B  & Aug13 & 1.6 & 3600 &  98   \\      
\object{A3556} F  & Aug13 & 1.3 & 7200 &  97    \\      
\object{A3558} B  & Jul14 & 1.5 & 3600 &  99   \\      
\object{A3558} F  & Aug13 & 1.5 & 7200 &  99    \\      
\object{A3560} B  & Jan15 & 1.8 & 3600 &  98\\
\object{A3560} F  & Jan15 & 1.6 & 7200 &  88 \\
\object{A3667} B  & Aug13 & 1.2 & 3600 &  98   \\      
\object{A3667} F  & Aug13 & 1.4 & 7200 &  96   \\      
\object{A3716} B  & Aug13 & 3.0 & 3600 &  95   \\      
\object{A3716} F  & Aug13 & 1.3 & 7200 &  94   \\        
\object{A3809} B  & Jul14 & 1.2 & 3600 &  100\\
\object{A3809} F  & Jul14 & 1.6 & 7200 &  100\\
\object{A3880} B  & Aug13 & 0.7 & 3600 &  98\\
\object{A3880} F  & Aug13 & 1.0 & 7200 &  96\\
\object{A4059} B  & Sep15 & 1.7 & 3600 & 99 \\
\object{A4059} F  & Sep15 & 1.8 & 7200 & 97 \\
\object{A500} B   & Dec13 & 1.8 & 5500 &  98 \\
\object{A500} F   & Dec13 & 1.6 & 7200 &  96\\
\object{A754} B   & Jan15 & 3.2 & 3600 &  89\\
\object{A754} F   & Dec13 & 1.4 & 7200 &  70\\
\object{A85} B    & Sep15 & 2.0 & 2400 & 99\\
\object{A957x} B  & Jan15 & 1.3 & 3600 &  96\\
\object{A970} B   & Dec13 & 3.5 & 3600 &  96\\
\object{A970} F   & Jan15 & 1.8 & 12000 & 90 \\
\object{IIZW108} B &Aug13 & 1.9 & 3600 &  99\\
\object{IIZW108} F &Sep15 & 1.4 & 7200 & 86 \\
\hline
\end{tabular} 
    \label{tab:configurations}
\end{table}

The data reduction was performed using the dedicated pipeline {\it 2dfdr \footnote{https://www.aao.gov.au/science/software/2dfdr}}, which properly handles science and calibration (flat--field and arc) exposures, producing wavelength-calibrated spectra for each arm.
After producing the single blue and red arm corrected spectra, we spliced them together using the region in common (at $\sim 5700$ \AA\,) in order to get a single spectrum covering the entire observed spectral range.
The spectra were corrected for the instrument sensitivity by the pipeline, while a proper flux calibration is missing.
For the purpose of the spectro-photometric analysis (that will be presented in a forthcoming paper), we performed a relative flux calibration using a set of spectra in common with the SDSS.
The sky subtraction was performed using the dedicated algorithm by \citet{sharp2010}, that implements the Principal Component Analysis to derive an optimal sky estimate starting from dedicated sky exposures.

\section{Data quality}\label{sec:quality}

\begin{figure}[h]
\begin{center}
\includegraphics[width=0.33\textwidth,angle=90]{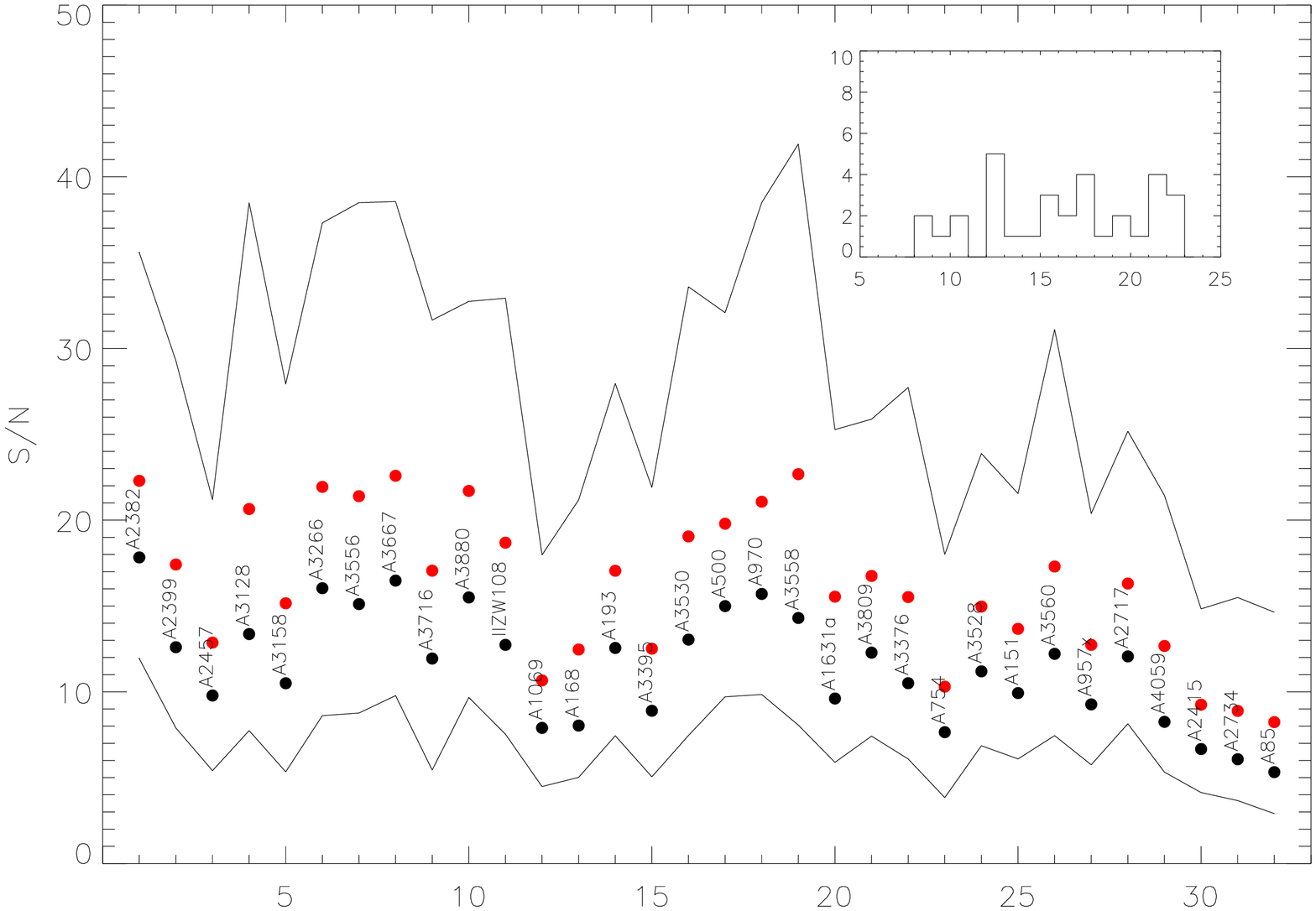}
\includegraphics[width=0.33\textwidth,angle=90]{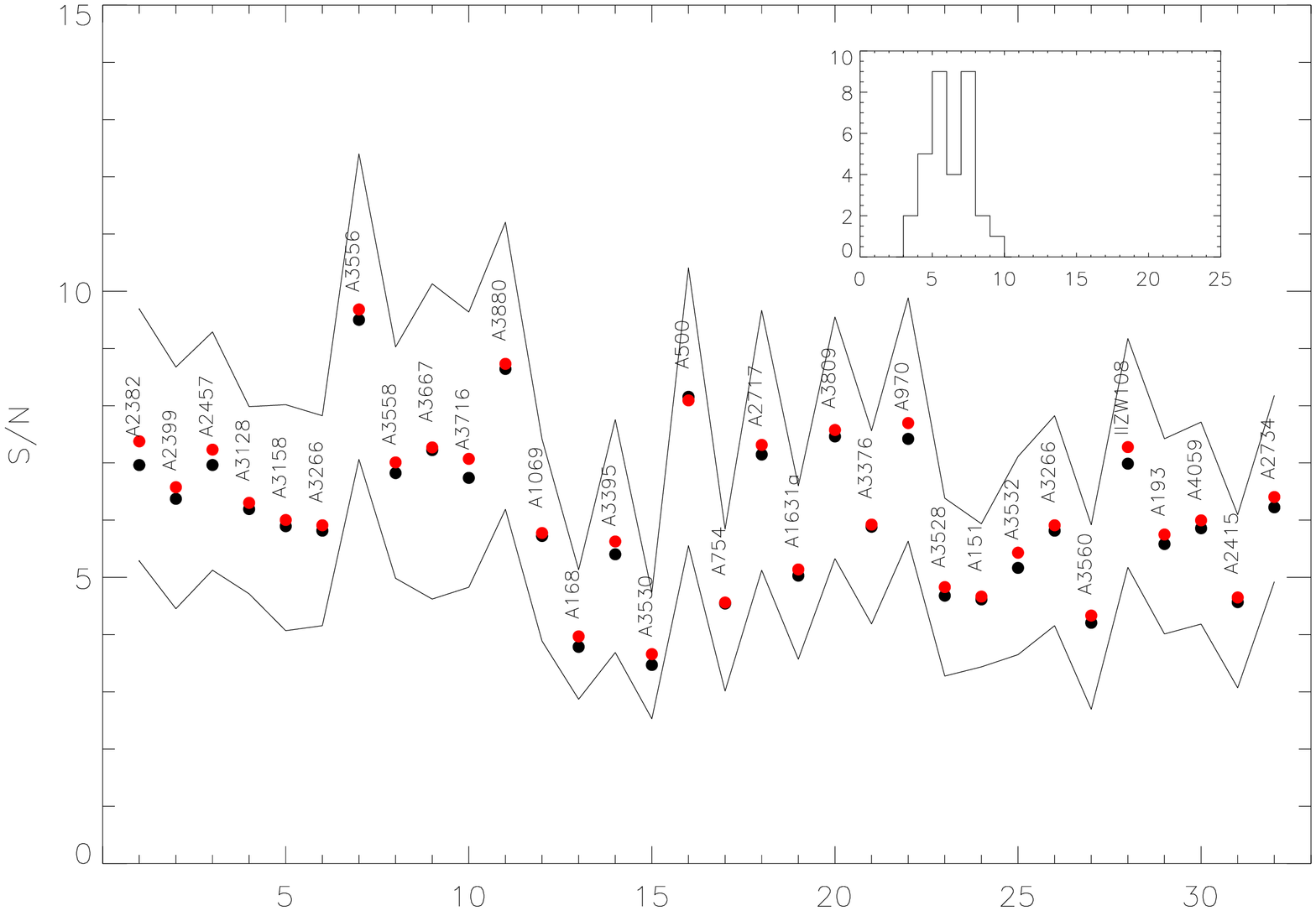}
\caption{Mean (red) and median (black) signal--to-noise together with the 68\% confidence limits for the {\it bright} (top panel) and {\it faint} (bottom panel) configurations.
In the insets, the average distributions of the measured signal--to--noise in the observed configurations (32 are bright and 32 are faint) are shown.
}
\label{fig:sn}
\end{center}
\end{figure}

We estimated the signal to noise ratios of the 17985 AAOmega spectra averaged over the entire spectrum from the root mean square (r.m.s.) of the spectrum itself \footnote{http://www.stecf.org/software/ASTROsoft/DER\_SNR}, obtaining values that go from approximately $ 11$ to $16$ (median and mean, respectively) for the bright configurations and from approximately $6$ to $6$ for the faint configurations.   Fig.\ref{fig:sn} shows the mean (in red) and the median (in black) signal to noise ratio together with the 68\% confidence limit for the observed configurations. 

The distributions of the average signal to noise (shown in the insets) reflect the objects magnitude range of the two configurations, with the bright configurations showing a larger range of measured S/N than the faint configurations.

\begin{figure*}[h]
\begin{center}
\includegraphics[width=0.63\textwidth,angle=90]{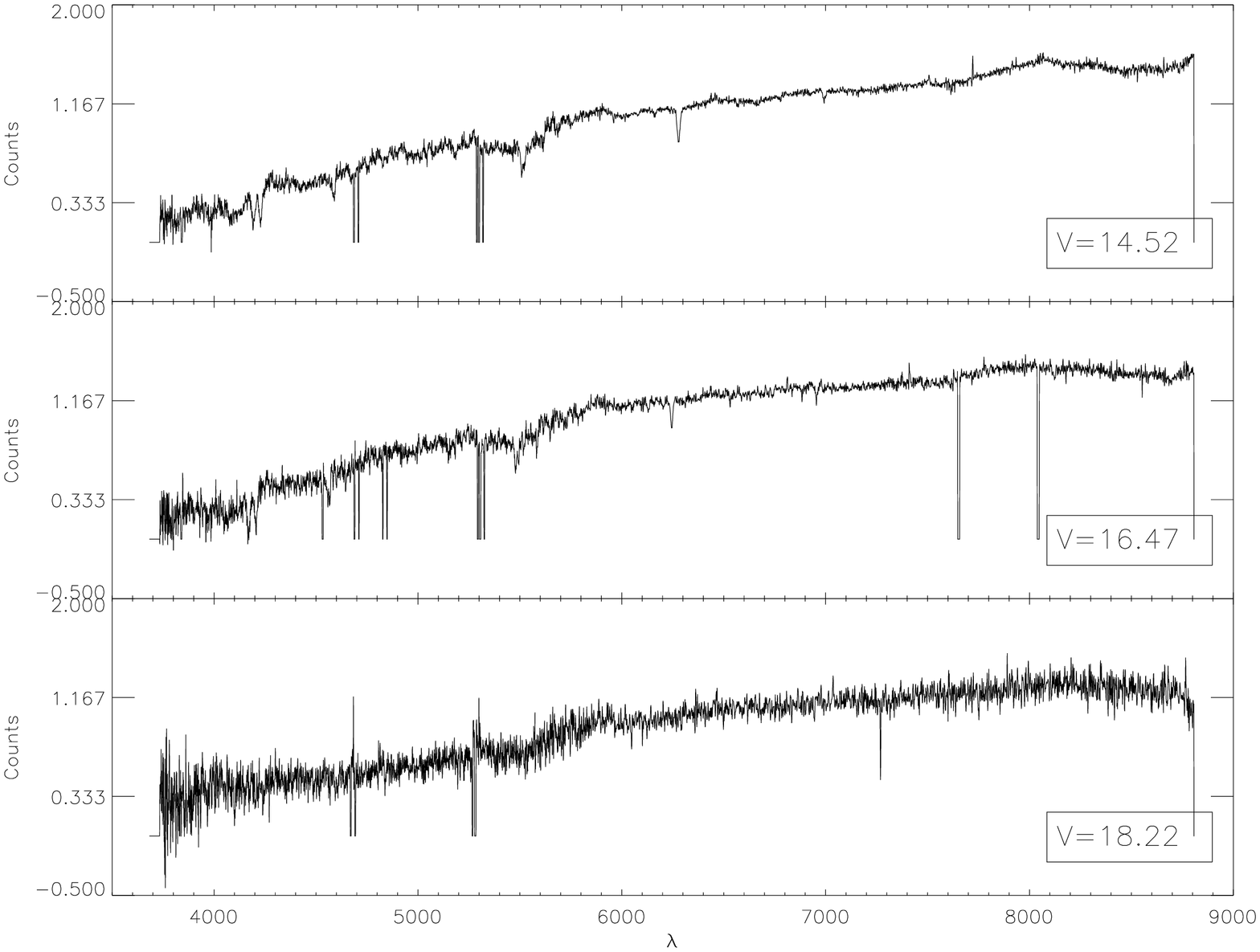}
\includegraphics[width=0.63\textwidth,angle=90]{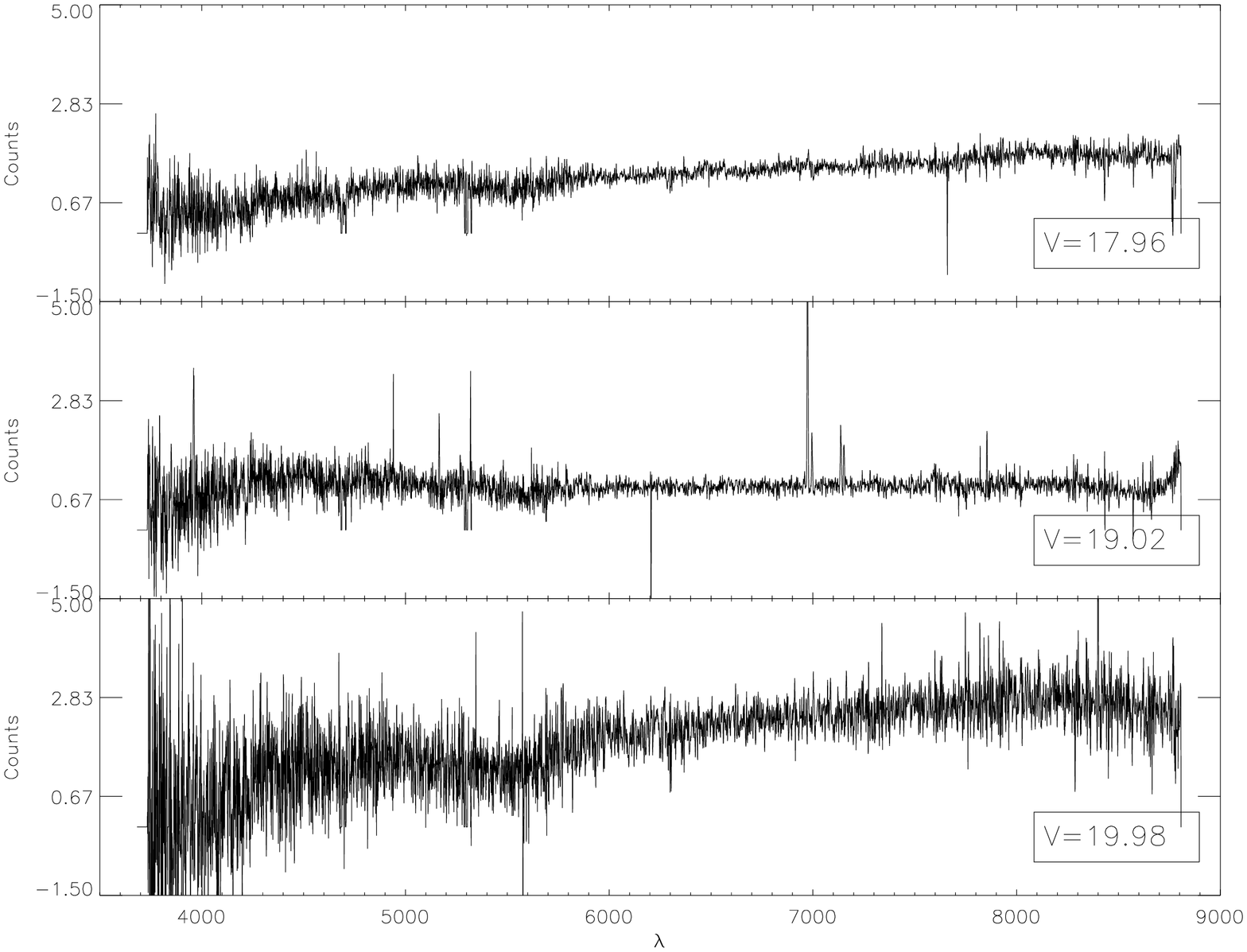}
\caption{Examples of observed spectra in the cluster \object{A2382}. Upper panel spectra belong to the {\it bright} configuration, while lower panel spectra belong to the {\it faint} configuration. {\it V} total magnitude is in the bottom right inset. The overall S/N is 55, 39, and 15 for the bright spectra, and 9, 6, and 4 for the faint ones.}\label{fig:spectra}
\end{center}
\end{figure*}

As an example,  in Fig.\ref{fig:spectra}, we show two sets of observed spectra, belonging to the {\it bright} (top panel) and {\it faint} (bottom panel) configurations of \object{A2382}.
The signal to noise ratio ranges from 55 to 15 in the brightest sources, while it goes down to 4 for the faintest ones. However, even for the faintest sources, we have been able to measure a redshift (see Sec.\ref{sec:redshifts}), mainly from emission lines.
The spectra shown are those covering the entire AAOmega spectral range, that is, the ones obtained by combining the red and the blue arm of the spectrograph. Due to the small overlap in the common region, the combination turns out to be non-optimal for the continuum match, even though it does not hamper the redshift determination. Three spectral regions are clearly affected by dead pixels ($[4660-4720$\AA],$[5258-5300$\AA],$[5560-5600$\AA]).

\section{Redshifts}\label{sec:redshifts}
We measured galaxy redshifts on the spliced spectra, that is, those extending over the entire wavelength range (3800 \AA-9000 \AA), in order to gather information both from the blue part of the spectra (where at our redshifts, most optical absorption lines are present) and from the red part of the spectra (where the $H_{\alpha}$+NII region is located).
The night sky residuals, present in specific regions of the spectra, have been replaced by interpolated values before measuring the redshifts.
On these cleaned spectra we measured both the redshift obtained by cross--correlating the spectrum with that of a template at zero velocity (using the {\it IRAF/xcsao} package) and the one obtained from the match of emission lines ({\it IRAF/emsao}).

Each single spectrum was visually inspected by members of the team (A. M., B. P., D. B., J. F., M. G., A. P., A. C.) to ensure that the automatic estimation was correct, and to assign a flag that was then used to derive the final redshift determination. We used;
\begin{itemize}
\item the weighted average of the two determinations if both were flagged as good, 
\item the emission line redshift if the cross-correlation was not convincing, and
\item the cross--correlation redshift if spectra did not contain emission lines (and the cross correlation was reliable).
\end{itemize}
Errors on the redshifts are the errors of the combined measurements of both absorption features and emission features if the two redshift estimates were averaged together, and are single errors on the measurements that do not have emission lines, or where we did not trust the absorption line cross-correlation.
Among the original 18995 spectra, we were able to measure 17985 redshifts, with a median error of 50 km/s and average completeness $\sim 95\%$.

\begin{figure}[h]
\begin{center}
\includegraphics[width=0.33\textwidth,angle=90]{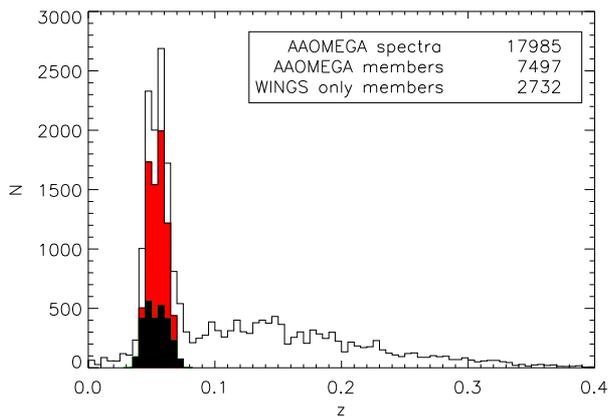}
\caption{Redshift distribution of OmegaWINGS targets: the empty histogram shows all redshifts and the red histogram shows the cluster members. The black histogram shows cluster members with a previous WINGS redshift that have not been re-observed with AAOmega.}
\label{fig:histo_tot}
\end{center}
\end{figure}
Figure \ref{fig:histo_tot} shows the distribution in redshift of all the OmegaWINGS targets (black line). The filled histogram represents the redshift determinations obtained with the previous WINGS spectroscopic survey \citep{Cava2009}, all in the inner cluster regions. The red histogram shows the new OmegaWINGS redshifts. When a new redshift was present, we used this determination instead of the old WINGS one (the difference between the two determinations being peaked at $0$, with a few cases of real mismatches in spectra with low S/N).
The number of cluster members with a AAOmega redshift is 7497, that adds to the 2732 galaxies belonging to the clusters for which we had a previous WINGS measurement. 
This means that the total number of cluster members rises to 10229 in total.

\subsection{Completeness}\label{sec:compl}
To ensure that we could properly derive global properties from the clusters under scrutiny, and from their galaxy populations, we calculated the spectroscopic completeness as a function of both magnitude and radial distance.
The first term $C_m$ is calculated as
\begin{equation}
C(m)=\frac{N_z(m)}{N_{ph}(m)}
,\end{equation}
and gives, as a function of magnitude, the number of targets with a measured redshift ($N_z(m)$) with respect to the number of target galaxies in the same magnitude bin present in the parent photometric catalog ($N_{ph}(m)$), that is, the photometric catalog from which we selected our spectroscopic target for the follow-up.

Recall that the candidates have to obey the following conditions:
\begin{itemize}
\item be classified as galaxies,
\item have a V total magnitude brighter than $V=20$ mag, and
\item have a $(B-V)$ color bluer than the one defined by the cluster red sequence.
\end{itemize}

The red sequence color varies slightly among clusters, and we took into account this effect in calculating the magnitude completeness.
As for the radial completeness, we calculated it as the ratio between the number of targets with redshift with respect to the number of candidates in the parent photometric catalog in the same radial bin 
, that is,
\begin{equation}
C(r)=\frac{N_z(r)}{N_{ph}(r)}
,\end{equation}
where radial bins have been chosen to have the same area.

\begin{figure}[htbp]
\begin{center}
\includegraphics[width=0.45\textwidth]{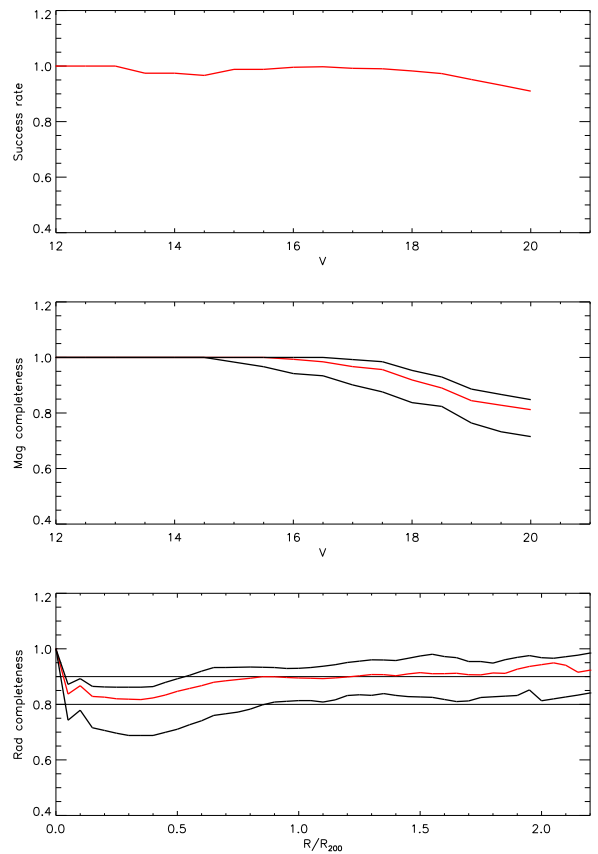}
\caption{Upper panel: success rate of redshift measurement. Middle panel: completeness in magnitude of the global OmegaWINGS spectroscopic sample. The red line gives the median value per bin of magnitude (in bins of 0.5 mag) and the two black lines represent the 15$^{th}$ and the 85$^{th}$ percentiles of the distribution. Lower panel: radial completeness.
}\label{fig:compl}
\end{center}
\end{figure}

Fig.\ref{fig:compl} shows, from top to bottom, the success rate of redshift measurements as a function of the V magnitude, the completeness in magnitude, and the radial completeness expressed in terms of $R/R_{200}$.

The upper panel of Fig.\ref{fig:compl} shows the success rate of the OmegaWINGS spectroscopic sample as a function of the total V magnitude of the targets, calculated as
\begin{equation}
S(m)=\frac{N_z(m)}{N_{targ}(m)}
,\end{equation}
and shows that at each magnitude bin, we were able to successfully measure redshifts in at least 90\% of targets.

While the completeness in magnitude (middle panel of Fig.\ref{fig:compl}) decreases at fainter luminosities, reaching a value of $80\%$ at V=20 mag, 
the radial completeness (lower panel of Fig.\ref{fig:compl}) is relatively flat above $0.5 R_{200}$, and is, on average, $90\%$. In the internal region, it is slightly lower ($80\%$) due to the possible fiber overlap and, mostly, to the lower priority given to targets already located inside the previous WINGS field of view. 

The bottom panel takes into account the fact that the clustercentric radial limit varies from cluster to cluster.

\section{Velocity dispersions and memberships}\label{sec:sigma}
The determination of cluster membership and of the cluster mean redshift is strictly related to the presence of possible interlopers in the cluster field of view. In our OmegaWINGS observations, this is even more problematic, given the very large field of view of our observations, that may be heavily contaminated by such objects in the external regions.
In order to give a reliable estimate of these parameters, we therefore adopted an iterative procedure, as in \citep{Cava2009}, following a $\pm3\sigma$ clipping algorithm, first described by \citet{Yahil1977}.

The procedure starts by eliminating candidates whose velocities are outside a given fixed range (in our case those with $z>|z_{cl}+0.015|$).
To this sample we apply an improved $3 \sigma$ clipping, which includes the weighted gap method \citep[e.g.,][]{Girardi1993} and interactively removes galaxies outside $R_{200}$. We proceed as follows.

Starting from the galaxies in the redshift range selected above and in the whole field of view with velocities given by $v = cz$, we apply the weighted gap method based on the ROSTAT routine \citep{Beers1990}, a widespread tool designed for robust estimation of simple statistics. Gaps are the measured gaps between the ordered velocities, here
defined as $g_i = v_i +1 - v_i$ for the sorted velocities, and weights as $w_i = i(N-i)$ for $i = 1,..,N-1$ for N
galaxies. The weighted gap is defined as 
\begin{equation}
\frac{\sqrt{gw}}{MM(\sqrt{gw})}
,\end{equation}
where MM is defined as the mean of the central $50\%$ of the data set and $w_i$ are a set of approximately Gaussian weights. As stated in \citet{Beers1991}, a weighted gap is considered significant if its value, relative to the mid mean of the
other weighted gaps formed from the same sample, is greater than 2.25.
In order to be conservative, we identified values larger than three as gaps, that thus suggest the possible presence of dynamical substructures altering the velocity dispersion determination. We visually inspected the velocity histograms to decide whether or not the gap found highlights the presence of a substructure and, if this is the case, galaxies identified as belonging to them have been removed from the calculation of the cluster velocity dispersion.
After the redefinition of structures, we computed the mean redshift $z_{cl}$ and the rest frame velocity dispersion $\sigma_{cl}$ for each cluster. 

The following steps have then been iterated until convergence in $\sigma_{cl}$ was reached: (1) Based on the ROSTAT routine \citep[see][]{Beers1990}, we used the biweight robust location and scale estimators to get $z_{cl}$ and $\sigma_{cl}$ and applied an iterative $3\sigma$ clipping until no rejection occurred. (2) We used the projected spatial distribution to define a further radius-dependent cut, that is, we determined $R_{200}$ and removed galaxies outside this radius.

Once the mean cluster redshifts and the velocity dispersions had been  determined, the memberships were derived for all galaxies, independently from the distances from the cluster center, here defined as the Brightest Cluster Galaxy (BCG).
The redshifts and velocity dispersions that are given in Tab.\ref{tab:sigma} were derived using only redshifts from the OmegaWINGS and WINGS surveys, therefore excluding additional redshifts present in the literature.
The errors quoted were obtained using the classical jackknife technique \citep{Efron1982}.

In four of our clusters, the procedure used to derive the membership suggested the presence of another structure close but separated in velocity, with a slightly different redshift (shown in Tab.\ref{tab:sigma}, last column).
More details will be given in Biviano et al. (in preparation) where we will analyze additional structures and the presence of substructures within the clusters with a dedicated algorithm.

\begin{figure*}[h]
\begin{center}
\includegraphics[width=0.85\textwidth]{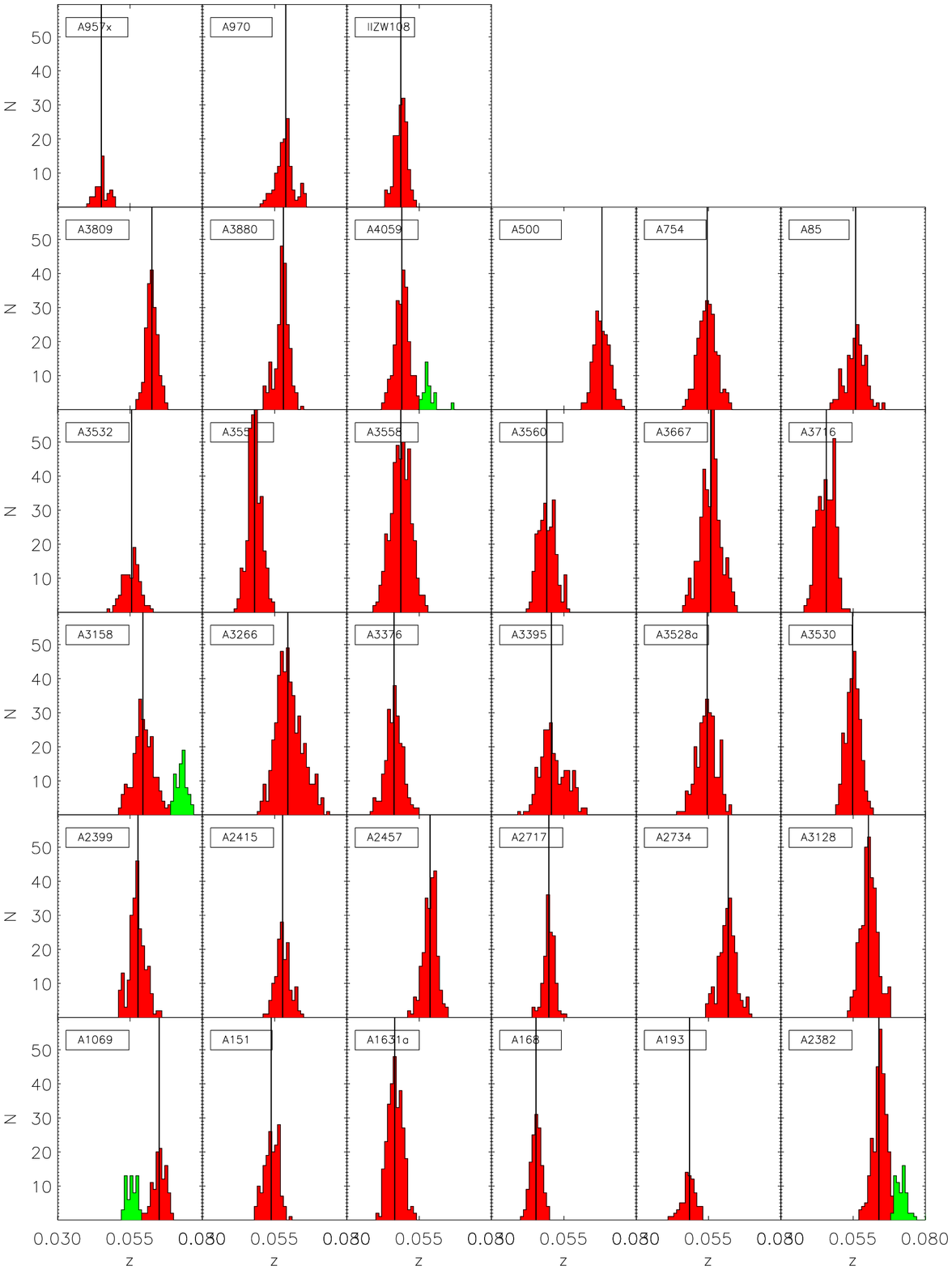}
\caption{Redshift distribution of the observed clusters as derived from the OmegaWINGS observations only. The vertical line shows the median cluster redshift (see sec.\ref{sec:sigma} for the details). Green histograms refer to other probable structures close in velocity.}
\label{fig:histoz}
\end{center}
\end{figure*}

Fig.\ref{fig:histoz} shows the redshift distribution for each cluster (in red), with the mean cluster redshift that we determined superimposed. Green histograms refer to galaxies belonging to the possible substructures.

In Tab.\ref{tab:sigma}, we give the number of targets observed with AAOmega with a reliable redshift determination (col. 2), the number of new cluster members (col. 3), the total number of members (including the ones present only in the WINGS spectroscopic catalog) that contributed to the velocity dispersion determination (col. 4), the velocity dispersion and its error in cols. 5 and 6, and the $R_{200}$ in Mpc.
The $R_{200}$ is the radius of the cluster region where the mean density is 200 times the critical density of the Universe, and has often been used as a proxy for the virial radius of a cluster.

Following \citet{Poggianti+2006}, we calculated $R_{200}$ as
\begin{equation}
R_{200} = 1.73 \frac{\sigma_{cl}}{1000 km s^{-1}} \frac{1}{\sqrt{\Omega_{\Lambda}+{\Omega_0}(1+z_{cl})^3}} h^{-1} Mpc
,\end{equation}
using the cluster velocity dispersion $\sigma_{cl}$ and the cluster redshift $z_{cl}$.

As a consequence, we derived the cluster masses (in $10^{15} M_{\odot}$) as in \citet{Poggianti+2006} using the relation between the cluster mass and the cluster velocity dispersion $\sigma_v$ and redshift $z_{cl}$ as in \citet{Finn+2005}, i.e.
\begin{equation}
M_{cl}=1.2 \left(\frac{\sigma_v}{1000 km s^{-1}}\right)^3 \times \frac{1}{\sqrt{\Omega_{\Lambda}+\Omega_0 (1+z_{cl})^3}}h_{100}^{-1} M_{\odot}
\end{equation}
having adopted $H_0=70$, $\Omega_{\Lambda}=0.7$ and $\Omega_0=0.3$.

\begin{figure*}[h]
\begin{center}
\includegraphics[width=0.85\textwidth]{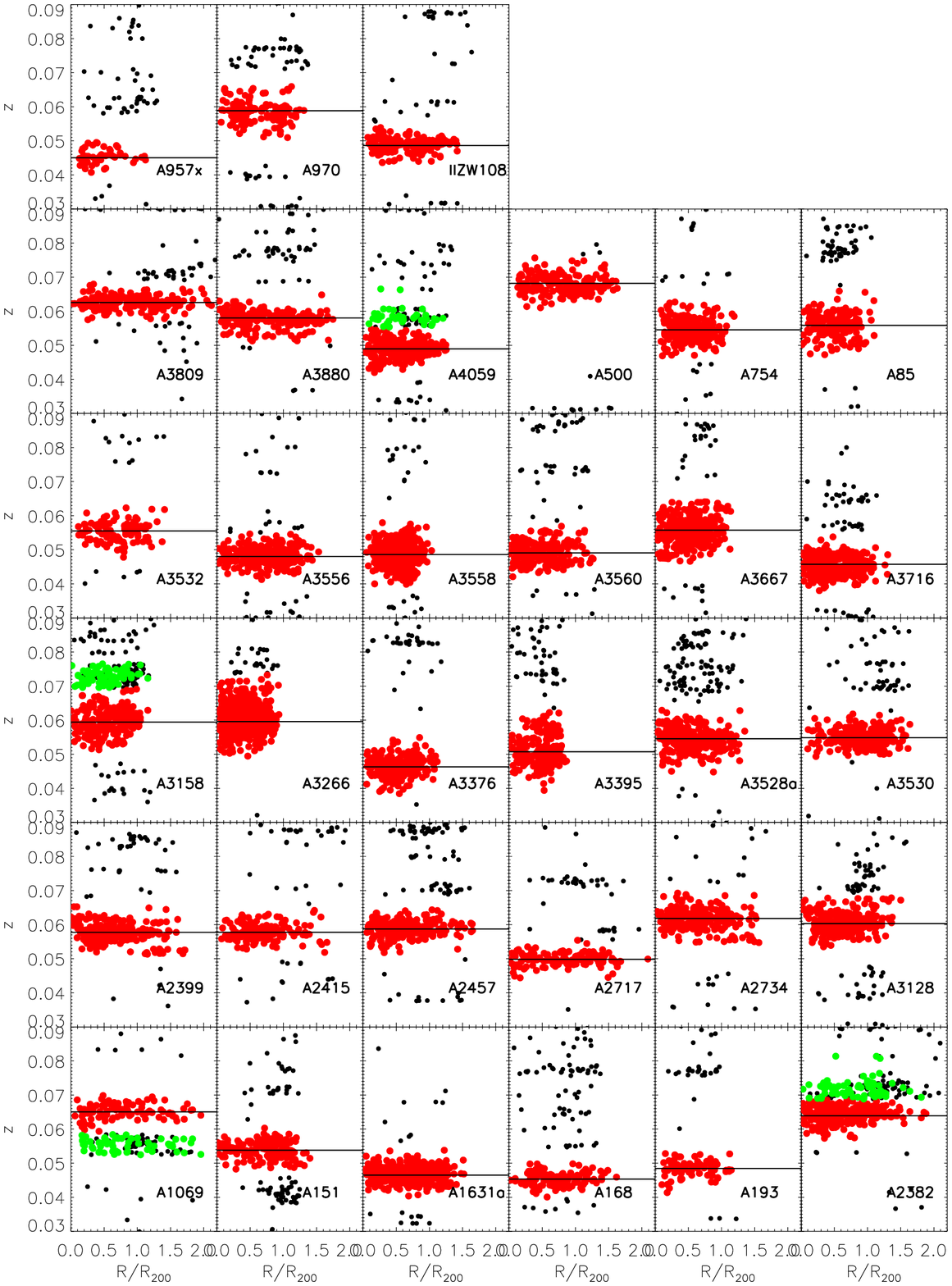}
\caption{Redshifts against distance from the cluster BCG, in terms of R/R$_{200}$. Red points show the cluster members, green points probable separate structures. Black dots are not members.
}
\label{fig:sigma}
\end{center}
\end{figure*}

Fig.\ref{fig:sigma} shows the redshifts as a function of the distance from the cluster BCG (expressed in terms of $R/R_{200}$) for the OmegaWINGS clusters: red dots are the cluster members, while black ones are foreground/background galaxies. The horizontal line indicates the cluster redshift derived using the procedure outlined above.
For all clusters the OmegaWINGS surveys allows to reach at least the virial radius, and for some of them even a much larger region (out to $2\times R_{200}$), where we still find cluster members.

\begin{table*}
\caption{OmegaWINGS main results}
\begin{center}
    \begin{tabular}{ lccccccccc}
\hline
Cluster  & N$_z$  & N$_{memb,OW}$& N$_{memb,tot}$  & z$_{cl}$ & $\sigma$  & $\sigma_{err}$ & R$_{200}$& M$_{200}$   & z$_{gap}$ \\
              &              &                            &                              &               & [km/s]       & [km/s]              & Mpc          & $10^{15} M_{\odot}$ & \\
\hline
\object{A1069}  &         496&         107&         130&  0.0651&         695&          54&  1.7& 0.56  & 0.058\\
\object{A151}   &         623&         165&         248&  0.0538&         738&          32&  1.8& 0.67  & 0.000\\
\object{A1631a} &         673&         288&         369&  0.0465&         760&          28&  1.8& 0.74  & 0.000\\
\object{A168}   &         459&         141&         141&  0.0453&         546&          38&  1.3& 0.27  & 0.000\\
\object{A193}   &         239&          72&         101&  0.0484&         764&          57&  1.8& 0.75  & 0.000\\
\object{A2382}  &         594&         271&         322&  0.0639&         698&          30&  1.7& 0.57  & 0.069\\
\object{A2399}  &         678&         234&         291&  0.0577&         729&          35&  1.7& 0.65  & 0.000\\
\object{A2415}  &         482&         144&         194&  0.0578&         690&          37&  1.7& 0.55  & 0.000\\
\object{A2457}  &         535&         232&         249&  0.0587&         679&          36&  1.6& 0.52  & 0.000\\
\object{A2717}  &         610&         135&         135&  0.0498&         544&          46&  1.3& 0.27  & 0.000\\
\object{A2734}  &         556&         220&         220&  0.0618&         780&          48&  1.9& 0.79  & 0.000\\
\object{A3128}  &         584&         333&         480&  0.0603&         838&          28&  2.0& 0.98  & 0.000\\
\object{A3158}  &         621&         243&         357&  0.0594&        1023&          37&  2.5& 1.79  & 0.069\\
\object{A3266}  &         877&         479&         678&  0.0596&        1318&          39&  3.2& 3.82  & 0.000\\
\object{A3376}  &         519&         229&         263&  0.0463&         844&          42&  2.0& 1.01  & 0.000\\
\object{A3395}  &         652&         244&         369&  0.0507&        1206&          55&  2.9& 2.94  & 0.000\\
\object{A3528}  &         627&         262&         262&  0.0545&        1016&          46&  2.4& 1.76  & 0.000\\
\object{A3530}  &         530&         275&         275&  0.0548&         674&          38&  1.6& 0.51  & 0.000\\
\object{A3532}  &         250&         107&         107&  0.0555&         805&          61&  1.9& 0.87  & 0.000\\
\object{A3556}  &         625&         328&         359&  0.0480&         668&          34&  1.6& 0.50  & 0.000\\
\object{A3558}  &         706&         442&         442&  0.0486&        1003&          33&  2.4& 1.69  & 0.000\\
\object{A3560}  &         580&         244&         283&  0.0491&         840&          35&  2.0& 0.99  & 0.000\\
\object{A3667}  &         687&         386&         386&  0.0558&        1010&          42&  2.4& 1.72  & 0.000\\
\object{A3716}  &         609&         327&         327&  0.0457&         848&          26&  2.0& 1.02  & 0.000\\
\object{A3809}  &         695&         189&         244&  0.0626&         553&          38&  1.3& 0.28  & 0.000\\
\object{A3880}  &         543&         216&         216&  0.0580&         688&          56&  1.7& 0.54  & 0.000\\
\object{A4059}  &         686&         229&         229&  0.0490&         752&          38&  1.8& 0.71  & 0.055\\
\object{A500}   &         478&         187&         227&  0.0682&         791&          43&  1.9& 0.82  & 0.000\\
\object{A754}   &         423&         250&         338&  0.0545&         919&          36&  2.2& 1.30  & 0.000\\
\object{A85}    &         359&         172&         172&  0.0559&         982&          55&  2.4& 1.58  & 0.000\\
\object{A957x}  &         154&          48&          92&  0.0451&         640&          47&  1.5& 0.44  & 0.000\\
\object{A970}   &         331&         136&         214&  0.0588&         844&          49&  2.0& 1.00  & 0.000\\
\object{IIZW108}&         504&         162&         171&  0.0486&         611&          38&  1.5& 0.38  & 0.000\\
\hline
\end{tabular}
\tablefoot{For each cluster in columns (1), (2), (3), and (4), we give the number of redshifts derived from OmegaWINGS spectroscopy, the number of OmegaWINGS members and the total number of members (including WINGS results), respectively. Column (5) gives the estimated mean cluster redshift, while columns (6) and (7) are the velocity dispersion and its error, as derived from the number of members listed in column (4). Column (8) and (9) are the $R_{200}$ in Mpc and the $M_{200}$ in units of $10^{15} M_{\odot}$, respectively, while in the last column we show the secondary redshift (see sec. \ref{sec:sigma} for details).}
\label{tab:sigma}
\end{center}
\end{table*}

Our estimates of cluster redshifts and velocity dispersions are in good agreement with those given by \citet{Cava2009}, except for A3395, where we measure a much larger velocity dispersion (1206 Km/s instead of 755 Km/s). This could be due to the larger extent of the OmegaWINGS fields, that reaches in this case $R_{200}$, while the WINGS spectroscopic survey just reached $\sim 0.5 R_{200}$.

\begin{figure*}[h]
\begin{center}
\includegraphics[width=0.85\textwidth]{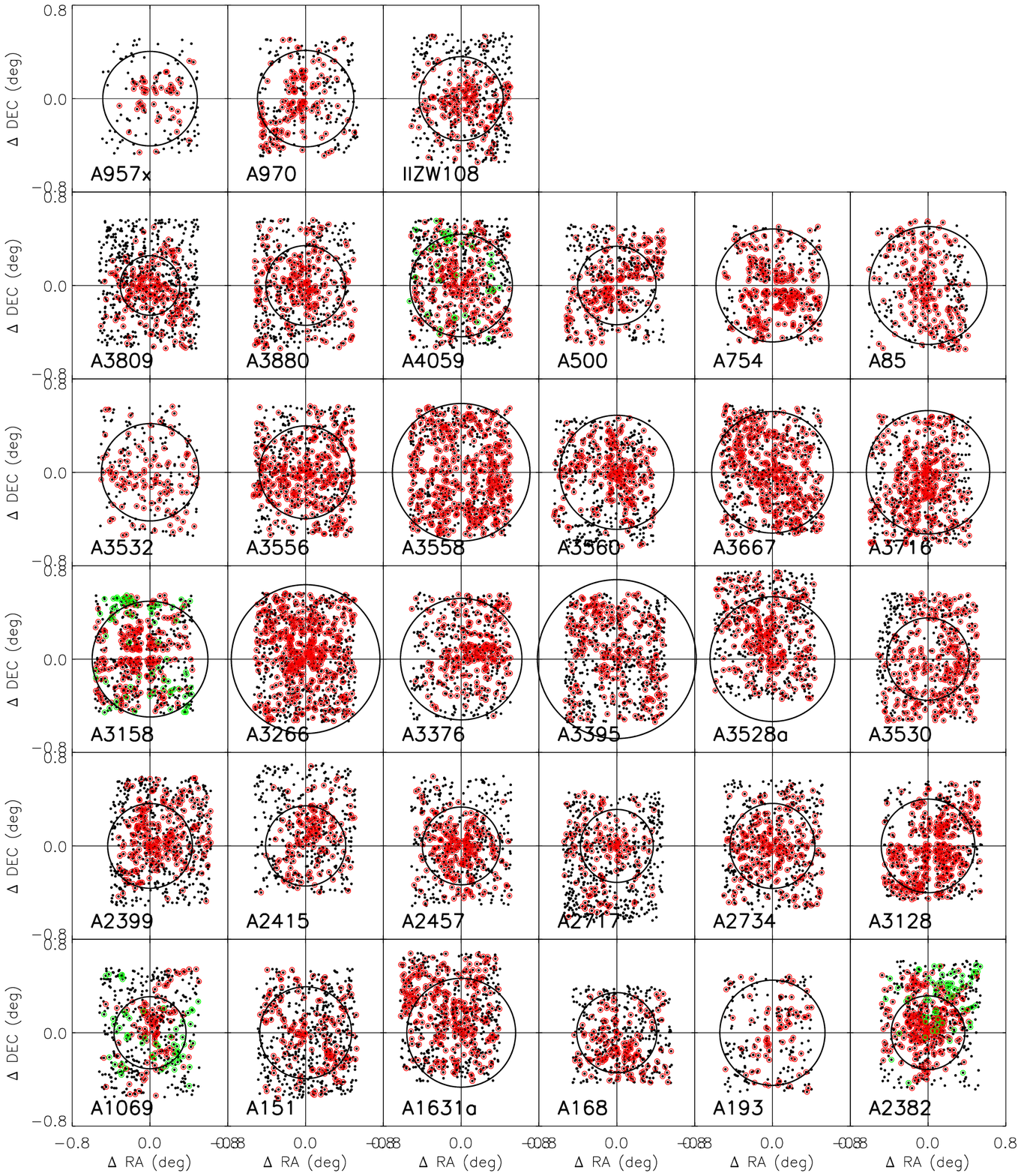}
\caption{Sky distribution of OmegaWINGS members (red symbols) and probable other structures (green symbols). Black dots are not members. The superimposed circle has a radius of $R_{200}$.
}
\label{fig:maps}
\end{center}
\end{figure*}

We show in Fig.\ref{fig:maps} the sky distributions of OmegaWINGS cluster members (red dots) together with the $R_{200}$ region (black circle), for each cluster.
The green circles are galaxies that might be related to the secondary structures identified by the analysis of the redshift distribution only.
\section{Access to the catalog}\label{sec:access}
The complete catalog of OmegaWINGS redshift will be available only at CDS and through the Virtual Observatory tool, as it is for the previous measurements derived from the WINGS/OmegaWINGS dataset \citep[see][for a complete description of the WINGS/OmegaWINGS database]{Moretti2014}.

The WINGS identifier has been inherited from the photometric parent catalog by \citet{Gullieuszik+2015}, and it is unique inside the WINGS/OmegaWINGS database.
We remind here that the identifier originates from the object coordinates, and that it has been newly assigned only when no match was present with sources already present in the original WINGS database (in order to avoid duplicates).

In the catalog, for each redshift, we give the OmegaWINGS identifier, the sky coordinates of the source, the redshift and its error, and the membership determined as previously described (flag 1 is for the galaxies considered cluster members, 0 for non-members).
Galaxies flagged with membership 2 are those belonging to the other structures defined here on the basis of a secondary peak in the redshift distribution.

\begin{table*}
\caption{OmegaWINGS redshifts}
\begin{center}
\small
    \begin{tabular}{ lcccccccc}
\hline
Cluster  & WINGS ID & RA (J2000) & DEC(J2000) & z & $z_{err}$  & M compl & R compl  & Memb \\
\hline
A1069   &    WINGSJ104034.89-083552.0   &  160.1454900   &  -8.5980600 & 0.17920  &  0.00064 & 0.788 & 0.789 & 0\\
A1069   &   WINGSJ104016.71-083547.4    &  160.0698700   &  -8.5968100 & 0.11306  &  0.00022 & 0.788 & 0.807 & 0\\
A1069   &  WINGSJ104105.27-083517.0     &  160.2719900   &  -8.5879700 & 0.05740  &  0.00016 & 0.788 & 0.842 & 2\\
A1069   &   WINGSJ104143.89-083206.8    &  160.4328900   &  -8.5352200 & 0.12495  &  0.00051 & 0.788 & 0.875 & 0\\
A1069   &    WINGSJ104042.94-083438.6   &  160.1788300   &  -8.5777400 & 0.09392  &  0.00046 & 0.788 & 0.797 & 0\\
A1069   &    WINGSJ104105.41-083212.3   &  160.2726900   &  -8.5367000 & 0.06503  &  0.00081 & 0.788 & 0.849 & 1\\
A1069   &    WINGSJ104027.03-083416.2   &  160.1127200   &  -8.5713900 & 0.09868  &  0.00012 & 0.788 & 0.789 & 0\\
A1069   &    WINGSJ104033.08-083146.9   &  160.1378900   &  -8.5299700 & 0.17782  &  0.00050 & 0.788 & 0.789 & 0\\
A1069   &    WINGSJ104123.59-082341.0   &  160.3482800   &  -8.3947100 & 0.21769  &  0.00019 & 0.788 & 0.875 & 0\\
A1069   &    WINGSJ104022.58-083037.8   &  160.0941500   &  -8.5107000 & 0.11511  &  0.00021 & 0.788 & 0.789 & 0\\
A1069   &    WINGSJ104017.22-083308.4   &  160.0718700   &  -8.5525100 & 0.11286  &  0.00042 & 0.788 & 0.789 & 0\\
A1069   &    WINGSJ104101.00-082438.1   &  160.2542400   &  -8.4106200 & 0.23828  &  0.00020 & 0.788 & 0.951 & 0\\
A1069   &    WINGSJ104130.86-081840.9   &  160.3786500   &  -8.3114000 & 0.26860  &  0.00022 & 0.788 & 0.778 & 0\\
A1069   &    WINGSJ104144.83-081515.6   &  160.4367800   &  -8.2543400 & 0.15127  &  0.00032 & 0.788 & 0.913 & 0\\
A1069   &    WINGSJ104106.03-082201.0   &  160.2751800   &  -8.3669600 & 0.31237  &  0.00046 & 0.788 & 0.900 & 0\\
A1069   &    WINGSJ104041.46-082648.1   &  160.1729400   &  -8.4469000 & 0.15228  &  0.00026 & 0.788 & 0.842 & 0\\
A1069   &    WINGSJ104026.93-082738.7   &  160.1123000   &  -8.4608300 & 0.31764  &  0.00025 & 0.788 & 0.797 & 0\\
A1069   &    WINGSJ104035.18-082746.3   &  160.1466100   &  -8.4630100 & 0.15894  &  0.00013 & 0.788 & 0.797 & 0\\
A1069   &    WINGSJ104138.64-080833.6   &  160.4110000   &  -8.1426700 & 0.06407  &  0.00008 & 0.788 & 1.000 & 1\\
A1069   &    WINGSJ104126.39-081046.8   &  160.3599500   &  -8.1796800 & 0.26871  &  0.00031 & 0.788 & 0.913 & 0\\
\end{tabular}
\tablefoot{Column (1) gives the cluster name, column (2) the unique OmegaWINGS/WINGS identifier, columns (3) and (4), the sky coordinates in degrees, column (5) and (6), the redshift  and the relative error, column (7) and (8), the completeness in magnitude and radial distance, respectively, and column (9), the membership. Cluster members are flagged 1 or 2, depending on the cluster structure/secondary structure, and 0 if they are not cluster members.}
\label{tab:redshifts}
\end{center}
\end{table*}
Radial and magnitude completeness were calculated as described in Sect. \ref{sec:compl}, and both must be taken into account when dealing with statistical studies, as we did in \citet{Vulcani+2011b,vulcani+2011c}.
OmegaWINGS spectra will be distributed in a forthcoming paper, where we will also give more details on the relative flux calibration and the spectrophotometric analysis.

\section{Summary}\label{sec:summary}
In this paper, we present the redshift determination for 17895 galaxies in z=0.04-0.07 clusters as derived from observations obtained with the AAOmega spectrograph at AAT.
Galaxies belong to 33 clusters out of the original 46 OmegaWINGS clusters observed with VST \citep{Gullieuszik+2015} over 1 square degree.

This sample of redshifts is aimed at giving a complete mapping of cluster galaxies out to large cluster--centric distances, that enables studies of galaxy transformations within clusters.

The high data quality expressed by the overall success rate of $95$\% in the redshift measurement, together with the radial and magnitude completeness ($90$\% and $80$\% at $V=20$ mag, respectively) characterize the sample as the most robust up to date for statistical studies of low--z cluster galaxies.

We have been able to determine new cluster redshifts and velocity dispersions, as well as cluster membership out to 1-2 virial radii depending on the cluster. 7497 galaxies turned out to be cluster members.
Cluster masses derived using our velocity dispersions range from $2.7\times10^{14} M_{\odot}$ to $3.8\times10^{15} M_{\odot}$ , therefore spanning a wide range in cluster properties.

\begin{acknowledgement}
Based on data acquired through the Australian Astronomical Observatory.
We acknowledge financial support from PRIN-INAF 2014.
B.V. acknowledges the support from an Australian Research Council Discovery Early Career Researcher Award (PD0028506).
This research has made use of the SIMBAD database, operated at CDS, Strasbourg, France.
We thank the anonymous referee for the helpful comments.
\end{acknowledgement}


\begin{thebibliography}{37}
\expandafter\ifx\csname natexlab\endcsname\relax\def\natexlab#1{#1}\fi

\bibitem[{Barkhouse {et~al.}(2007)Barkhouse, Yee, \&
  L{\'{o}}pez-Cruz}]{Barkhouse07}
Barkhouse, W.~A., Yee, H. K.~C., \& L{\'{o}}pez-Cruz, O. 2007, The
  Astrophysical Journal, 671, 16

\bibitem[{Beers {et~al.}(1990)Beers, Flynn, \& Gebhardt}]{Beers1990}
Beers, T.~C., Flynn, K., \& Gebhardt, K. 1990, The Astronomical Journal, 100,
  32

\bibitem[{Beers {et~al.}(1991)Beers, Forman, Huchra, Jones, \&
  Gebhardt}]{Beers1991}
Beers, T.~C., Forman, W., Huchra, J.~P., Jones, C., \& Gebhardt, K. 1991, The
  Astronomical Journal, 102

\bibitem[{Cava {et~al.}(2009)Cava, Bettoni, Poggianti, Couch, Moles, Varela,
  Biviano, DOnofrio, Dressler, Fasano, Fritz, Kjaergaard, Ramella, \&
  Valentinuzzi}]{Cava2009}
Cava, A., Bettoni, D., Poggianti, B.~M., {et~al.} 2009, Astronomy and
  Astrophysics, Volume 495, Issue 3, 2009, pp.707-719, 495, 707

\bibitem[{Dressler(1980)}]{dressler80}
Dressler, A. 1980, The Astrophysical Journal, 236, 351

\bibitem[{Efron(1982)}]{Efron1982}
Efron, B. 1982, CBMS-NSF Regional Conference Series in Applied Mathematics,
  Philadelphia: Society for Industrial and Applied Mathematics (SIAM), 1982

\bibitem[{Fasano {et~al.}(2006)Fasano, Marmo, Varela, D'Onofrio, Poggianti,
  Moles, Pignatelli, Bettoni, Kj{\ae}rgaard, Rizzi, Couch, \&
  Dressler}]{Fasano+2006}
Fasano, G., Marmo, C., Varela, J., {et~al.} 2006, Astronomy {\&} Astrophysics,
  445, 805

\bibitem[{Fasano {et~al.}(2015)Fasano, Poggianti, Bettoni, Donofrio, Dressler,
  Vulcani, Moretti, Gullieuszik, Fritz, Omizzolo, Cava, Couch, Ramella, \&
  Biviano}]{Fasano2015}
Fasano, G., Poggianti, B.~M., Bettoni, D., {et~al.} 2015, Monthly Notices of
  the Royal Astronomical Society, Volume 449, Issue 4, p.3927-3944, 449, 3927

\bibitem[{Finn {et~al.}(2005)Finn, Zaritsky, McCarthy, Poggianti, Rudnick,
  Halliday, Milvang?Jensen, Pello, \& Simard}]{Finn+2005}
Finn, R.~A., Zaritsky, D., McCarthy, Jr., D.~W., {et~al.} 2005, The
  Astrophysical Journal, 630, 206

\bibitem[{Girardi {et~al.}(1993)Girardi, Biviano, Giuricin, Mardirossian, \&
  Mezzetti}]{Girardi1993}
Girardi, M., Biviano, A., Giuricin, G., Mardirossian, F., \& Mezzetti, M. 1993,
  The Astrophysical Journal, 404, 38

\bibitem[{Gomez {et~al.}(2003)Gomez, Nichol, Miller, Balogh, Goto, Zabludoff,
  Romer, Bernardi, Sheth, Hopkins, Castander, Connolly, Schneider, Brinkmann,
  Lamb, SubbaRao, \& York}]{Gomez2002}
Gomez, P., Nichol, R., Miller, C., {et~al.} 2003, The Astrophysical Journal,
  Volume 584, Issue 1, pp. 210-227., 584, 210

\bibitem[{Gullieuszik {et~al.}(2015)Gullieuszik, Poggianti, Fasano, Zaggia,
  Paccagnella, Moretti, Bettoni, D'Onofrio, Couch, Vulcani, Fritz, Omizzolo,
  Baruffolo, Schipani, Capaccioli, \& Varela}]{Gullieuszik+2015}
Gullieuszik, M., Poggianti, B., Fasano, G., {et~al.} 2015, Astronomy {\&}
  Astrophysics, Volume 581, id.A41, 17 pp., 581

\bibitem[{Haines {et~al.}(2011)Haines, Merluzzi, Busarello, Dopita, Smith,
  La~Barbera, Gargiulo, Raychaudhury, \& Smith}]{Haines2011}
Haines, C.~P., Merluzzi, P., Busarello, G., {et~al.} 2011, Monthly Notices of
  the Royal Astronomical Society, Volume 417, Issue 4, pp. 2831-2845., 417,
  2831

\bibitem[{Jaff{\'{e}} {et~al.}(2011)Jaff{\'{e}}, Arag{\'{o}}n-Salamanca,
  Kuntschner, Bamford, Hoyos, De~Lucia, Halliday, Milvang-Jensen, Poggianti,
  Rudnick, Saglia, Sanchez-Blazquez, \& Zaritsky}]{jaffe2011}
Jaff{\'{e}}, Y.~L., Arag{\'{o}}n-Salamanca, A., Kuntschner, H., {et~al.} 2011,
  Monthly Notices of the Royal Astronomical Society, Volume 417, Issue 3, pp.
  1996-2019., 417, 1996

\bibitem[{Lewis {et~al.}(2002)Lewis, Balogh, De~Propris, Couch, Bower, Offer,
  Bland-Hawthorn, Baldry, Baugh, Bridges, Cannon, Cole, Colless, Collins,
  Cross, Dalton, Driver, Efstathiou, Ellis, Frenk, Glazebrook, Hawkins,
  Jackson, Lahav, Lumsden, Maddox, Madgwick, Norberg, Peacock, Percival,
  Peterson, Sutherland, \& Taylor}]{Lewis2002}
Lewis, I., Balogh, M., De~Propris, R., {et~al.} 2002, Monthly Notices of the
  Royal Astronomical Society, Volume 334, Issue 3, pp. 673-683., 334, 673

\bibitem[{Merluzzi {et~al.}(2014)Merluzzi, Busarello, Haines, Mercurio, Okabe,
  Pimbblet, Dopita, Grado, Limatola, Bourdin, Mazzotta, Capaccioli, Napolitano,
  \& Schipani}]{Merluzzi2015}
Merluzzi, P., Busarello, G., Haines, C.~P., {et~al.} 2014, Monthly Notices of
  the Royal Astronomical Society, Volume 446, Issue 1, p.803-822, 446, 803

\bibitem[{Merluzzi {et~al.}(2010)Merluzzi, Mercurio, Haines, Smith, Busarello,
  \& Lucey}]{Merluzzi2010}
Merluzzi, P., Mercurio, A., Haines, C.~P., {et~al.} 2010, Monthly Notices of
  the Royal Astronomical Society, 402, 753

\bibitem[{Miszalski {et~al.}(2006)Miszalski, Shortridge, Saunders, Parker, \&
  Croom}]{aao_configure}
Miszalski, B., Shortridge, K., Saunders, W., Parker, Q.~A., \& Croom, S.~M.
  2006, Monthly Notices of the Royal Astronomical Society, Volume 371, Issue 4,
  pp. 1537-1549., 371, 1537

\bibitem[{Moran {et~al.}(2007)Moran, Miller, Treu, Ellis, \& Smith}]{Moran2007}
Moran, S.~M., Miller, N., Treu, T., Ellis, R.~S., \& Smith, G.~P. 2007, The
  Astrophysical Journal, Volume 659, Issue 2, pp. 1138-1152., 659, 1138

\bibitem[{Moretti {et~al.}(2015)Moretti, Bettoni, Poggianti, Fasano, Varela,
  D'Onofrio, Vulcani, Cava, Fritz, Couch, Moles, \&
  Kj{\ae}rgaard}]{Moretti2015}
Moretti, A., Bettoni, D., Poggianti, B.~M., {et~al.} 2015, Astronomy {\&}
  Astrophysics, Volume 581, id.A11, 49 pp., 581

\bibitem[{Moretti {et~al.}(2014)Moretti, Poggianti, Fasano, Bettoni,
  DÕOnofrio, Fritz, Cava, Varela, Vulcani, Gullieuszik, Couch, Omizzolo,
  Valentinuzzi, Dressler, Moles, Kj{\ae}rgaard, Smareglia, \&
  Molinaro}]{Moretti2014}
Moretti, A., Poggianti, B.~M., Fasano, G., {et~al.} 2014, Astronomy {\&}
  Astrophysics, 564, A138

\bibitem[{Paccagnella {et~al.}(2016)Paccagnella, Vulcani, Poggianti, Moretti,
  Fritz, Gullieuszik, Couch, Bettoni, Cava, Fasano, \&
  D'Onofrio}]{Paccagnella2016}
Paccagnella, A., Vulcani, B., Poggianti, B.~M., {et~al.} 2016, The
  Astrophysical Journal Letters, Volume 816, Issue 2, article id. L25, 6 pp.
  (2016)., 816

\bibitem[{Pimbblet {et~al.}(2001)Pimbblet, Smail, Kodama, Couch, Edge,
  Zabludoff, \& O'Hely}]{Pimbblet2002}
Pimbblet, K.~A., Smail, I., Kodama, T., {et~al.} 2001, Monthly Notices of the
  Royal Astronomical Society, Volume 331, Issue 2, pp. 333-350., 331, 333

\bibitem[{Poggianti {et~al.}(2006)Poggianti, von~der Linden, De~Lucia, Desai,
  Simard, Halliday, Aragon?Salamanca, Bower, Varela, Best, Clowe, Dalcanton,
  Jablonka, Milvang?Jensen, Pello, Rudnick, Saglia, White, \&
  Zaritsky}]{Poggianti+2006}
Poggianti, B.~M., von~der Linden, A., De~Lucia, G., {et~al.} 2006, The
  Astrophysical Journal, 642, 188

\bibitem[{Popesso {et~al.}(2006)Popesso, Biviano, B{\"{o}}hringer, \&
  Romaniello}]{Popesso06}
Popesso, P., Biviano, A., B{\"{o}}hringer, H., \& Romaniello, M. 2006,
  Astronomy and Astrophysics, 445, 29

\bibitem[{Sharp \& Parkinson(2010)}]{sharp2010}
Sharp, R. \& Parkinson, H. 2010, Monthly Notices of the Royal Astronomical
  Society, Volume 408, Issue 4, pp. 2495-2510., 408, 2495

\bibitem[{Sharp {et~al.}(2006)Sharp, Saunder, Smith, Churilov, Correll, Dawson,
  Farrel, Frost, Haynes, Heald, Lankshear, Mayfield, Waller, \&
  Whittard}]{Sharp2006}
Sharp, R., Saunder, W., Smith, G., {et~al.} 2006, Ground-based and Airborne
  Instrumentation for Astronomy. Edited by McLean, Ian S.; Iye, Masanori.
  Proceedings of the SPIE, Volume 6269, id. 62690G (2006)., 6269

\bibitem[{Smith {et~al.}(2004)Smith, Saunders, Bridges, Churilov, Lankshear,
  Dawson, Correll, Waller, Haynes, \& Frost}]{Smith2004}
Smith, G.~A., Saunders, W., Bridges, T., {et~al.} 2004, in Ground-based
  Instrumentation for Astronomy. Edited by Alan F. M. Moorwood and Iye
  Masanori. Proceedings of the SPIE, Volume 5492, pp. 410-420 (2004)., Vol.
  5492, 410

\bibitem[{Smith {et~al.}(2012)Smith, Lucey, Price, Hudson, \&
  Phillipps}]{Smith2012}
Smith, R.~J., Lucey, J.~R., Price, J., Hudson, M.~J., \& Phillipps, S. 2012,
  Monthly Notices of the Royal Astronomical Society, 419, 3167

\bibitem[{Valentinuzzi {et~al.}(2010)Valentinuzzi, Fritz, Poggianti, Cava,
  Bettoni, Fasano, D'Onofrio, Couch, Dressler, Moles, Moretti, Omizzolo,
  Kj{\ae}rgaard, Vanzella, \& Varela}]{Valentinuzzi+2010}
Valentinuzzi, T., Fritz, J., Poggianti, B.~M., {et~al.} 2010, The Astrophysical
  Journal, 712, 226

\bibitem[{Valentinuzzi {et~al.}(2009)Valentinuzzi, Woods, Fasano, Riello,
  D'Onofrio, Varela, Bettoni, Cava, Couch, Dressler, Fritz, Moles, Omizzolo,
  Poggianti, \& Kj{\ae}rgaard}]{Valentinuzzi2009}
Valentinuzzi, T., Woods, D., Fasano, G., {et~al.} 2009, Astronomy {\&}
  Astrophysics, 501, 851

\bibitem[{Vulcani {et~al.}(2011{\natexlab{a}})Vulcani, Poggianti,
  Arag{\'{o}}n-Salamanca, Fasano, Rudnick, Valentinuzzi, Dressler, Bettoni,
  Cava, D'Onofrio, Fritz, Moretti, Omizzolo, \& Varela}]{Vulcani+2011b}
Vulcani, B., Poggianti, B.~M., Arag{\'{o}}n-Salamanca, A., {et~al.}
  2011{\natexlab{a}}, Monthly Notices of the Royal Astronomical Society, 412,
  246

\bibitem[{Vulcani {et~al.}(2011{\natexlab{b}})Vulcani, Poggianti, Dressler,
  Fasano, Valentinuzzi, Couch, Moretti, Simard, Desai, Bettoni, DÕOnofrio,
  Cava, \& Varela}]{vulcani+2011c}
Vulcani, B., Poggianti, B.~M., Dressler, A., {et~al.} 2011{\natexlab{b}},
  Monthly Notices of the Royal Astronomical Society, 413, 921

\bibitem[{Vulcani {et~al.}(2012)Vulcani, Poggianti, Fasano, Desai, Dressler,
  Oemler, Calvi, DÕOnofrio, \& Moretti}]{Vulcani2012}
Vulcani, B., Poggianti, B.~M., Fasano, G., {et~al.} 2012, Monthly Notices of
  the Royal Astronomical Society, 420, 1481

\bibitem[{Wenger {et~al.}(2000)Wenger, Ochsenbein, Egret, Dubois, Bonnarel,
  Borde, Genova, Jasniewicz, Lalo{\"{e}}, Lesteven, \& Monier}]{simbad}
Wenger, M., Ochsenbein, F., Egret, D., {et~al.} 2000, Astron. Astrophys. Suppl.
  Ser, 143, 9

\bibitem[{Yahil \& Vidal(1977)}]{Yahil1977}
Yahil, A. \& Vidal, N.~V. 1977, The Astrophysical Journal, 214, 347

\bibitem[{York(2000)}]{sdss}
York, D.~G. 2000, The Astronomical Journal, 120, 9

\end{thebibliography}
\end{document}